\documentclass[amsmath,amssymb,10pt,floatfix,two column,preprintnumbers] {revtex4}
\usepackage[dvips]{graphicx}
\begin{document}
\title{\bf Dynamics of Fluctuation Dominated Phase Ordering: Hard-core 
Passive Sliders on a Fluctuating Surface} 
\author{Sakuntala Chatterjee and Mustansir Barma}
\affiliation{Department of Theoretical Physics, Tata Institute of Fundamental
Research, Homi Bhabha Road, Mumbai-400005, India}
\preprint{TIFR/TH/05-40}
\begin{abstract}
We study the dynamics of a system of 
 hard-core particles sliding downwards on a
one dimensional fluctuating interface,
which in a special case can be mapped to the problem of a passive scalar
 advected
by a Burgers fluid. Driven by the surface fluctuations, the particles show 
a tendency to cluster, but the hard-core interaction prevents collapse. 
We use numerical simulations to measure the auto-correlation function
in steady state and in the aging regime, and  
space-time correlation functions in steady state.
We have also calculated these quantities analytically in a related 
surface model.  The steady state auto-correlation is a 
scaling function of $t/L^z$, where $L$ is the system size and $z$ the 
dynamic exponent. Starting from a finite intercept, the scaling function
decays with a cusp, in the small argument limit. The finite value of
the intercept indicates the existence of long range order in the system.
 The space-time correlation,
which is a function of $r/L$ and $t/L^z$, is  non-monotonic
in $t$ for fixed $r$. The aging 
auto-correlation is a scaling function of $t_1$ and $t_2$ where
 $t_1$ is the waiting time and $t_2$ 
the time difference. This scaling function 
 decays as a power law for $t_2 \gg  t_1$; for
$t_1 \gg  t_2$, it decays with a cusp as in steady state. 
To reconcile the occurrence of strong fluctuations in the steady state
with the fact of an ordered state,  
we measured the distribution function of the 
length of the largest cluster. This shows
that fluctuations never destroy ordering, but rather the
system meanders from one ordered configuration to another on a relatively rapid 
time scale. 
\end{abstract}
\maketitle
\section{Introduction}

The concentration of a substance advected by a driving field such as a 
fluid flow often shows
interesting behavior. In examples such as smoke dispersed in air or
fluorescent dye carried by a turbulent jet, an initial
local concentration of the passive particles generally spreads out in 
space~\cite{siggia}.
However, if the fluid is 
compressible, instead of spreading out, the advected substance may show
a tendency to cluster, as with air
bubbles in water or dust particles in air~\cite{falkovich,vergassola}.
\par
 In the situations discussed above, the
 advected  substance has a negligible effect on the fluid flow. These 
 problems are examples of `passive scalar'
 problems, where the dynamics of a non-equilibrium driving field strongly
 affects that of the
 other (passive) field with no back-effect from the latter. In this paper,  
 we study clustering in a set of passive particles which are subjected to a 
fluctuating force field. The specific system we study consists of hard-core 
particles sliding under gravity on a one dimensional fluctuating interface; the 
instantaneous force on a particle is then
 proportional to the local slope of the surface. Interestingly, our results
also pertain to particles advected by a fluid, using the fact that 
the equation governing a moving interface 
can be mapped onto the Burgers equation, which describes a compressible 
fluid~\cite{medina}.
\par
The degree of clustering of the particles depends strongly
on the interactions between them.
Two cases have been studied earlier---particles which are completely 
non-interacting~\cite{nagar,drossel} and particles which interact via hard-core
 exclusion~\cite{das,manoj}. We study the latter, more realistic, case 
 in this paper and address two broad issues, namely dynamics and ordering.
\par
 The first issue concerns the time-dependent
properties of the passive particles. We obtain results for the 
 auto-correlation and 
space-time correlation functions in steady
state, and aging correlations in the approach to steady state. To our knowledge,
the dynamics of passive scalars has not been explored systematically, and our
study adds to the relatively sparse work on this important
 question~\cite{mitra}. We have considered two kinds of surface evolutions---
namely those that respect symmetry under reflection (Edwards-Wilkinson type) 
and those which break this symmetry (Kardar-Parisi-Zhang type).
Our simulation results for the dynamics of the sliding particle system
 are supplemented by analytical calculations for a 
related model of coarse-grained interface variables.
\par
 The second issue concerns the characterization of the steady state as an
ordered state. Earlier studies of static properties revealed that
strong fluctuations are  present in the steady state and do not decrease 
 even in the thermodynamic limit~\cite{das}. On the other hand, the
scaling function for the density correlation indicates 
that the system has  long range order.   
 This sort of fluctuation dominated phase
ordering (FDPO) is characterized by a broad distribution of the order parameter.
The question then arises: 
In what sense does FDPO represent an ordered 
state, if strong fluctuations drive it between 
 macroscopically different configurations on a relatively rapid time-scale?
 We address this by studying the variation
of the length of the 
largest particle cluster present in the system and show that the corresponding 
probability distribution provides an unequivocal signal of ordering.

\section{Overview}
\subsection{Surface Fluctuation and Particle Movement}
A surface with no overhangs is completely specified by the 
height $ h(x,t) $ at point $x$ at time $t$. 
The evolution of the height field is taken to be described by the 
Kardar-Parisi-Zhang (KPZ) equation~\cite{medina}.
\begin{equation}
\frac{\partial h}{\partial t} = \nu_1 \frac{\partial^2 h}{\partial x^2}
+ \lambda \left ( \frac{\partial h}{\partial x} \right ) ^2 +  \eta_1 (x,t) 
\end{equation}  
The first term represents the smoothening effect of surface
 tension $\nu_1$, and  $\eta_1 (x,t)$ is a white noise
 with zero average and $ \langle \eta_1 (x,t) \eta_1 (x',t' ) \rangle = \Gamma 
\delta (x-x') \delta (t-t')$. Notice that if $\lambda \neq 0$,
  $h \rightarrow -h$ symmetry is not preserved, reflecting the fact that  the
 interface moves in a preferred direction. 
However, if $\lambda =0$, the equation has an
 $h \rightarrow -h$ symmetry and describes the
 Edwards-Wilkinson (EW) model~\cite{ew}. 
\par
  The height-height correlation function
has a scaling form for large separations in space
and time~\cite{barabasi} :
\begin{equation}
\langle [h(x,t)-h(x',t')]^2 \rangle \approx |x-x'|^{2\chi } f \left ( 
\frac{|t-t'|}{|x-x'|^z} \right )  
\label{eq:kpz}
\end{equation} 
Here $f$ is a scaling function and $\chi$ and $z$ are the roughness and
 dynamic exponents, respectively, with values which depend on the 
surface dynamics.
For an EW interface $\chi = 1/2$, $z=2$ while for a KPZ interface $\chi = 1/2$,
$z=3/2$.
\par
The hard-core particles slide downwards along the local slope $\left (
\frac{\partial h}{\partial x} \right )$ of the interface. In the overdamped
 limit, their velocity is proportional to the local gradient of height.
The equation governing the evolution of particle density can be derived
from  the continuity equation 
$\frac{\partial \rho (x,t)}{\partial t} = -\frac{\partial J (x,t)}{\partial x}$.
The local current $J(x,t)$ has a systematic part $\rho (1-\rho) 
(1-2\frac{\partial h}{\partial x})$, a 
diffusive part $-\nu_2 \frac{\partial \rho}{\partial x}$ (driven by local
density inhomogeneity) and a stochastic part $\eta_2 (x,t)$ (a Gaussian white 
noise).  The time-evolution equation for the density fluctuation 
$\tilde{\rho}=\rho - \rho_0$ is then
\begin{eqnarray}
\nonumber
\frac{\partial \tilde{\rho}}{\partial t}\!\!\!\!\!\! & 
\!\!\!\!\!\! = \nu_2 \frac{\partial ^2 \tilde{\rho}}
{\partial x^2 } +2\rho_0 (1-\rho_0 )\frac{\partial ^2 h}{\partial x^2 } \\
\nonumber
&  - (1-2\rho_0 -2 \tilde{\rho}) \left ( \frac{\partial \tilde{\rho}}
{\partial x} \right ) \left [1-2\left ( \frac{\partial h}{\partial x} \right )
\right ] +\\
&  2(1-2\rho_0) \tilde{\rho} \frac{\partial ^2 h}{\partial x^2 } - 
2\tilde{\rho}^2  \frac{\partial ^2 h}{\partial x^2 }
+ \frac{\partial \eta_2 (x,t)}{\partial x}
\label{eq:sp}
\end{eqnarray}
We will not analyze this equation directly; rather we will study the
particle dynamics by performing numerical simulations on a lattice model,
described in section 3, 
whose long distance and long time properties are expected to be described
by Eqs.(\ref{eq:kpz}) and (\ref{eq:sp}).

\subsection{FDPO : Static Properties}
 In an earlier study on static properties 
of this model~\cite{das}, the density-density 
correlation $C(r,L)$ of the sliding particles was
 measured and found to be a scaling function
 of $r/L$ in the scaling limit $r \rightarrow \infty$, $L \rightarrow \infty$
with $r/L$ fixed, as for phase-ordered states.
  The scaling function has a finite intercept $m^2$, and for small argument
it decays with a cusp:
\begin{eqnarray}
C(r,L) &=& f \left ( \frac{r}{ L} \right ) \label{eq:f}\\
&=& {m}^2 \left ( 1-a\left (\frac{r}{ L} \right )
 ^{\alpha} \right ),
 \; \; \left ( \frac{r}{L} \ll 1 \right ) \label{eq:alpha}
\end{eqnarray}
$m^2$ is a measure of long range order (LRO) as LRO is defined 
by the large $r$ (scaling limit) behavior of the correlation function. In
the limit of 
an infinite system, this corresponds to $r/L \rightarrow 0$. 
\par
The value of the cusp exponent $\alpha$ depends on the dynamics of the
driving surface: $\alpha \simeq 0.5$ for 
particles on an EW interface while $\alpha \simeq 0.25$ when the
 underlying surface 
is of KPZ type~\cite{das}. In a related coarse-grained surface model
defined in~\cite{das} and discussed in section $3$ below, 
the correlation function was shown analytically to have the above scaling
form with $m^2=1$ and $\alpha=0.5$ for both EW and KPZ surfaces.
For hard-core particles sliding on a two dimensional surface, the same scaling form is
found for $C(r,L)$, but with a different value of the intercept and the cusp
exponent.
For customary phase-ordering systems, it is expected that this scaling function 
should decay linearly, consistent with the Porod law ~\cite{porod}.
 The cusp ($\alpha < 1 $) is one manifestation of the unusual nature of FDPO. 
 
\par
For a phase-ordered system, 
the lowest
non-zero Fourier component of the density profile
 measures the extent of phase-separation and is an
appropriate  order parameter. A characteristic of FDPO is that
the distribution of this order parameter remains
broad even in the thermodynamic limit, indicating the presence of strong 
fluctuations. 

\subsection{FDPO : Dynamical Properties}
In this paper, we study the dynamics associated with FDPO. Our results are
summarized below. We find that   
 the steady state auto-correlation
in the density fluctuation $\sigma(x,t)$ of the sliding particles (SP):  
$A_{SP}(t) = \langle \sigma(x,0) \sigma(x,t) \rangle $ is a 
scaling function of $t/L^z$, consistent with the notion of phase ordering.
 For $t \ll L^z$, this scaling function decays with
a cusp, in contrast to a linear decay normally
expected for phase-ordering systems.  
The presence of the cusp is the dynamical manifestation of the 
unconventional character of FDPO. 
\par
We also monitored 
the auto-correlation function in the aging regime 
${\cal A}_{SP}(t_1 , t_2) = \langle \sigma(x,t_1) \sigma(x,t_1+t_2) \rangle $. 
From the dynamic scaling hypothesis,  ${\cal A}_{SP}(t_1 , t_2)$
 should be
 a scaling function of $t_1/t_2$, the ratio of initial time to the time 
lag~\cite{a.bray}. In the limit $t_2 \gg t_1$, the 
scaling function shows a power law decay 
 with an exponent that depends on the 
phase-ordering kinetics. In the opposite limit $t_1 \gg t_2$, the scaling 
function decays with a cusp as in steady state, the only difference being 
that the system size $L$, as appears in the steady state scaling function, 
is replaced by the coarsening length $t_1^{1/z}$.  
\par
The space-time correlation for the sliding particles
in steady state, defined as $G_{SP}(r,t) = \langle 
\sigma (x,0) \sigma (x+r,t) \rangle$  
 is a function of $r/L$ and $t/L^z$. When
 plotted against $t/L^z$ for a fixed value of $r/L$, this function
is non-monotonic.
and decays like the 
steady state auto-correlation for large $t$.
 The auto-correlation function in steady state 
and in the aging regime, together with the space-time correlation function in 
steady
state constitute our dynamical characterization of FDPO.
   
\subsection{FDPO : An Ordered State?}
The distinctive feature of FDPO is the presence
of strong fluctuations in steady state. On the one hand, a diagnostic of
LRO such as a non-zero value of the intercept $m^2$ indicates an ordered state.
On the other hand, the macroscopic state of the system changes relatively 
 rapidly over a time-scale $\sim L^z$, in contrast to a normal
 phase-ordered system, where the
typical time  grows exponentially with $L$. The small lifetime of a given 
macroscopic ordered state gives pause, and seems to contradict the notion
of LRO. To resolve this 
apparent contradiction we study the time dependence of the 
length of the largest particle cluster $l_{max}(t) $ .
We conclude that the short lifetime is associated with 
 the system wandering over a multitude of ordered 
states, each very different from the other, but all 
characterized by a large value of $l_{max}(t)$. Dynamical excursions away from
this attractor of ordered states are extremely infrequent and associated with
an exponentially growing time scale.

\section{Description of the Model }
We study a discrete model of a  fluctuating 
interface on which hard-core particles slide downwards under gravity, 
following the local slope of the interface. 
The $1$-d interface of length $L$, consists of discrete surface
 elements; the slope of the surface elements between the $i$-th and $(i+1)$-th 
site is $\tau_{i+\frac{1}{2}}$, which can take the value $+1$ or $-1$. 
Accordingly the height at site $i$ is given by $h_i=\sum_{j=1}^{i}
 \tau_{j-\frac{1}{2}}$. The dynamics of the interface involves stochastic
 corner flips with exchange of adjacent $\tau$'s; the transition $/ \setminus$
 to $\setminus /$ occurs with a rate $p_1$ while  $\setminus /$  to 
$/ \setminus$ with rate $q_1$. We have taken $p_1=q_1=1$ for an EW surface and
$p_1=1$, $q_1=0$ for KPZ surface.
The overall slope ${\cal T} =\frac{1}{L}
\sum_{i=1}^{L}
\tau_{i+\frac{1}{2}} $ is conserved and in our case we will consider
 ${\cal T} =0$, meaning that the interface is untilted.
\par
The hard-core particles are represented by  variables $\{\sigma_i\}$ 
each of which 
takes a value $+1$ or $-1$ according as the $i$-th site contains a particle
or a hole. The deviation from half-filling
 ${\cal S}=\frac{1}{L}\sum_{i=1}^{L}\sigma_{i}$ is  
conserved and we consider ${\cal S}=0$, corresponding to the case of
 half-filling.
A particle and hole on adjacent sites $(i,i+1)$ exchange with
rates that depend on the intervening local slope $\tau_{i+\frac{1}{2}}$; 
thus the moves $\bullet \backslash \circ $ $\rightarrow$ 
$\circ \backslash \bullet $ and $\circ / \bullet $ $\rightarrow$
  $\bullet / \circ $ occurs at rate $p_2$ while the inverse moves occur at
 rate $q_2$. In the case when the particles are sliding downwards 
along gravity, we have  $q_2 < p_2 $. We have considered $p_2=1$ and $q_2=0$.
Because of the hard-core exclusion 
between the
particles, with the above update rules, the system has a particle-hole 
symmetry, i.e. any correlation function involving the density variable
remains invariant as the particle density is replaced by the hole density.
This implies that 
unlike the case of non-interacting particles where the correlation 
function of the density variable is qualitatively different between the 
advection case and the anti-advection case~\cite{nagar}, the two processes show
identical correlation behavior here. 
\par
From the dynamical rules, it follows that the movement of particles depends
on the fluctuations of the underlying interface. Due to gravity the particles  
 tend to slide down into local valleys. However, in the
non-equilibrium system under consideration,
 before the particles can fill in the lowest valley, the
 interface evolves, often causing the valley to turn over. 
Nevertheless, it is useful to consider the adiabatic 
limit where the interface moves infinitely more slowly than the particles, 
in which case the 
particles have ample time to explore the landscape and eventually
 settle in the deepest valleys. 
It seems plausible that the dynamics of hills and valleys of
the interface may provide insight into the 
 dynamics of the particles. This motivates the definition of 
a coarse-grained depth model (CD model) as follows~\cite{das}. Consider the 
variable $s_i(t)$ defined as $s_i(t)=sgn [h_i(t)-\langle h(t) \rangle ]$, 
where $\langle h(t) \rangle $ is the average height at time $t$: $\langle h(t)
 \rangle=\frac{1}{L} \sum_{i=1}^{L} h_i(t) $. The variable $s_i(t)$ 
can take values 
$+1$,$-1$ or $0$, depending on whether the position of the $i$-th site
is above, below or at the average level. In other words, $s_i(t)$ gives 
a coarse-grained description of the surface by labeling `highlands' and
`lowlands'.
For an EW interface, the dynamics is tractable and we obtain an analytic
 expression 
for time-dependent correlations of $s_i(t)$. These results might be expected
to be close to those of $\sigma_i (t)$ in the extreme adiabatic limit.
 As a matter
of fact, we find that they also describe qualitatively the particle model
 even in the strongly non-equilibrium case. 

\section{Auto-correlation function in steady state}
We have studied the auto-correlation $A(t,L)$
 involving the density variable
 $\langle \sigma_i (0) \sigma_i (t) \rangle$  in the sliding particle (SP)
model and also that for the CD variables  
 $\langle s_i (0) s_i (t) \rangle $ in the CD model. Periodic boundary
conditions are used. We 
 will see below that in the steady state
of a system of size $L$, the auto-correlation  $A(t,L)$
 is a scaling function of $\frac{t}{L^z}$, where $z$ is the dynamic exponent
  defined earlier. This scaling function shows a cusp in the small 
argument limit, as seen previously in the static correlation scaling
 function [Eq. (\ref{eq:alpha})]:
\begin{eqnarray}
A(t,L) &=& h\left ( \frac{t}{L^z} \right ) \label{eq:h}\\
&=& {m}^2 \left [ 1-b \left ( \frac{t}{L^z} \right ) ^{\beta '} \right ], \; \; \;  \frac{t}{L^z} \rightarrow 0   \label{eq:beta}
\end{eqnarray}
 $m$ is a measure of the LRO as 
explained in section 2. Note that for large enough time, the auto-correlation 
function is expected to factorize and become equal to $m^2$.
\par
However, for small time,  $t \lesssim 1$, which falls outside the scaling 
regime, 
 the auto-correlation function 
shows a linear drop with an $L$-dependent slope:
\begin{eqnarray}
A(t,L) \approx 1-b' \frac{t}{L^{\lambda}}, \hspace{1cm}
 \left ( t \lesssim 1 \right )   
\label{eq:lambda}
\end{eqnarray}
If $m^2=1$, as shown below for the  
 CD model,  matching
Eqs.(\ref{eq:beta}) and (\ref{eq:lambda}) for $t \simeq 1$ yields 
\begin{equation}
\lambda = z \beta'
\label{reln}
\end{equation} 
If $m^2 \neq 1$, as happens 
for the SP model, a relation between the exponents 
cannot be obtained. Instead, the matching condition determines
 a time scale $t^*$
for the crossover from the linear decay in Eq.(\ref{eq:lambda})
to the cuspy decay in Eq.(\ref{eq:beta}). 
In the large $L$ limit, we find to the leading order, 
\begin{equation}
t^* = \frac{1-m^2}{b'} L^{\lambda}
\end{equation}
\begin{table}
\begin{tabular}{|c||c|c||c|c|}
\hline
&\multicolumn{2}{c|}{CD Model}&\multicolumn{2}{c|}{SP Model}\\
\cline{2-5}
&EW & KPZ & EW & KPZ\\
\hline
\hline
$m$ & $1.0$ & $1.0$ & $0.82\pm0.03$ & $0.75\pm0.04$\\
\hline
$\beta '$ & $0.25$ & $0.31\pm0.02 $ & $0.22\pm0.02$ & $0.18\pm0.01$\\
\hline
$\lambda$ & $0.5$ & $0.5$ & $0.26\pm0.005$ & $0.15\pm0.005$\\
\hline
$\gamma$ & $0.75$ & $0.84\pm0.02$ & $0.69\pm0.02$ & $0.82\pm0.04$\\
\hline
\end{tabular}
\caption{\it The values of relevant exponents and order parameter for dynamical
characterization of CD model and SP model}
\end{table}  
\par
Let us illustrate these properties, by discussing
the auto-correlation 
 in the CD model, defined as $A_{CD}(t,L)=\langle s_i (0) s_i (t) \rangle $. 
First consider  short
times $t \lesssim 1$. At $t=0$ let the initial configuration of the surface
be $\{ h_i (0) \} $. As time passes, there are stochastic
corner flips, as described in section $3$. However, only those flips occurring
close to the average level can cause a change in the CD variable $s_i (t)$,
 as any
local fluctuation far above or below the average level, would
not change the sign of $s_i (t)=(h_i (t) - \langle h(t) \rangle ) $.  
 More precisely, only those sites in $\{ h_i (0) \} $ which have at least one
 neighbor situated exactly on the average level, putatively contribute 
to the drop in auto-correlation function.
 Now, for a self-affine surface of length $L$ and 
 roughness exponent $\chi $, the number of such points scales as $L^{1-\chi }$
and the density of such points goes as $L^{-\chi }$~\cite{meakin}.
 For small $t$, 
the probability
that any one of these points will actually take part in a local fluctuation
is proportional to $t$. This immediately implies 
$A_{CD}(t \lesssim 1,L) \approx 1-{b_1}' \frac{t}{L^{\chi}}$. 
Comparison with Eq.(\ref{eq:lambda}) shows that for the CD
 model, we have $\lambda=\chi=\frac{1}{2} $. Note that although EW and KPZ
 surfaces have
 different dynamics, the above argument holds for both of them as their 
stationary measure is the same in $1$-$d$. 
\begin{figure} [h]
\includegraphics[scale=0.7,angle=0]{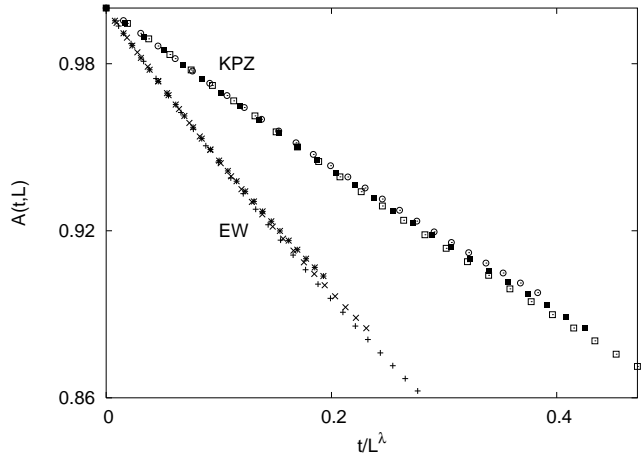}
\caption{\it Illustrating the linear drop of $A(t,L)$ for short times 
$t \lesssim 1$ in the SP model for system size $L=128,256,512$.}
\label{fig:micsp}
\end{figure}
\par
For the particle model, 
although the initial drop is found to be linear as described in 
Eq.(\ref{eq:lambda}), the exponent $\lambda$ takes the value $0.26\pm0.005$ for
 particles on an EW surface and $0.15\pm0.005$ for 
particles on a KPZ surface. The data are shown in Fig.(\ref{fig:micsp}).  
\par
For $t \geq 1$, we have analytically calculated $A_{CD}(t,L)$ for 
an EW 
interface. This exploits the fact that $h_i(t)$ in this 
case is a Gaussian variable, implying $s_i$ correlations satisfy the 
 arc-sine law  
\begin{equation}
\langle s_i(t)s_i(0) \rangle = \frac{2}{\pi } \sin ^{-1} \left ( \frac{\langle 
H_i(t)H_i(0) \rangle }{\sqrt {\langle {H_i}^2 (t) \rangle \langle {H_i}
^2 (0) \rangle }} \right )
\end{equation}
where $H_i(t)=h_i(t)-\langle h(t) \rangle$, which is also a Gaussian variable.
If $\tilde {h}_k(t)$ is the Fourier transform of $h_i(t)$, the numerator
 in the argument of arcsine can be written as 
$\sum_{k \neq 0} \langle \tilde{h}_k(t)\tilde{h}_{-k}(0) \rangle = 
  \sum_{k \neq 0} \Gamma \frac{\exp \left ( -c_{k} t \right ) }
{c_k }$, using the discrete version of the EW equation. Here, 
$c_k = 4\nu_1 \sin ^2 \frac{k}{2}$.
 Moreover, $\langle  {H_i}^2 (t) \rangle 
= \langle  {H_i}^2 (0) \rangle 
= \Gamma \sum_{k \neq 0 } \frac{1}{c_k}$.  
Thus we have 
\begin{equation}
\langle s_i(t)s_i(0) \rangle = \frac{2}{\pi} \sin ^{-1} \left [ \frac
{\sum_{k \neq 0} \frac{\exp \left ( -c_k t \right )}{c_k}}{\sum_{k \neq 0}
 \frac{1}{c_k}} \right ]
\label{eq:stsum}
\end{equation}
We have numerically evaluated this discrete sum and plotted it in
 Fig.(2a) against the scaling argument $t/L^2$ for different $L$
values. The cusp exponent can be read off from the plot in
the inset.
\par
 In the continuum limit, Eq.(\ref{eq:stsum}) becomes 
\begin{equation}
\langle s(x,t)s(x,0)\rangle = \frac{2}{\pi} \sin ^{-1} \left [ \frac
{\int_{\frac{2\pi }{L}}^{\pi }dk \frac{\exp \left (-k^2 t \right ) }{k^2}} 
{\int_{\frac{2\pi }{L}}^{\pi } \frac{dk}{k^2}} \right ]
\end{equation}
The integral in the numerator takes the form  
$\frac{L\Gamma}{\pi}\left [\frac{L}{2\pi}\exp \left ( -\frac{4
\pi^2 t}{L^2 } \right ) +
 \sqrt{\pi t}\; erf \left ( \frac{2 \pi \sqrt{t}}{L} \right ) -\sqrt{
\pi t}\right ]$.  
In the limit $t/L^2 \ll 1$, this becomes, to the leading order, $\frac{L\Gamma}
{\pi}\left [\frac{L}{2\pi}- \sqrt{\pi t} \right ]$. Noting
that  the denominator is
 $\frac{L\Gamma}{\pi}. \frac{L}{2\pi}$ and expanding for small values of 
$\frac{\sqrt{t}}{L}$, we get
\begin{equation}
\langle s(x,t)s(x,0)\rangle \approx 1-\frac{4}{\pi^{\frac{1}{4}}} \left (
 \frac{t}{L^2} \right )^{1/4} \hspace{0.5cm} \left ( t/L^2 \ll 1 \right )  
\end{equation}
Comparing with Eq.(\ref{eq:beta}) gives $m^2=1$, $\beta ' = \frac{1}{4}$,
 $z=2$. 
\par
For the KPZ surface, the time evolution equation for the height field is not
Gaussian and hence such an analytical
treatment is not possible.
 We study $A_{CD}(t)$ using Monte Carlo simulation. No initial
equilibration is required as 
the steady state measure for a KPZ surface with periodic boundary conditions
gives equal weight to every configuration. 
We followed the update rules discussed in section 2 and 
averaged over sites  
as well as over $10^5$ histories. The results
are shown in Fig.(2b). A good scaling collapse
is obtained for different $L$, on rescaling the time to $t/L^z$ with       
 $z=\frac{3}{2}$. The cusp exponent $\beta '$ was extracted
by plotting $m^2-A_{CD}(t)$ against $t/L^z$ (shown in the inset), using
$m^2=1$ and this gives $\beta '$ to be $0.31\pm0.02 $. Our best estimate
corresponds to the largest system size $L=2048$. 
The error-bar is based on the values of $\beta '$ obtained  
 for smaller system size ($L=512,1024$); the statistical error is much smaller.
\begin{figure} [h!]
\includegraphics[scale=0.3,angle=270]{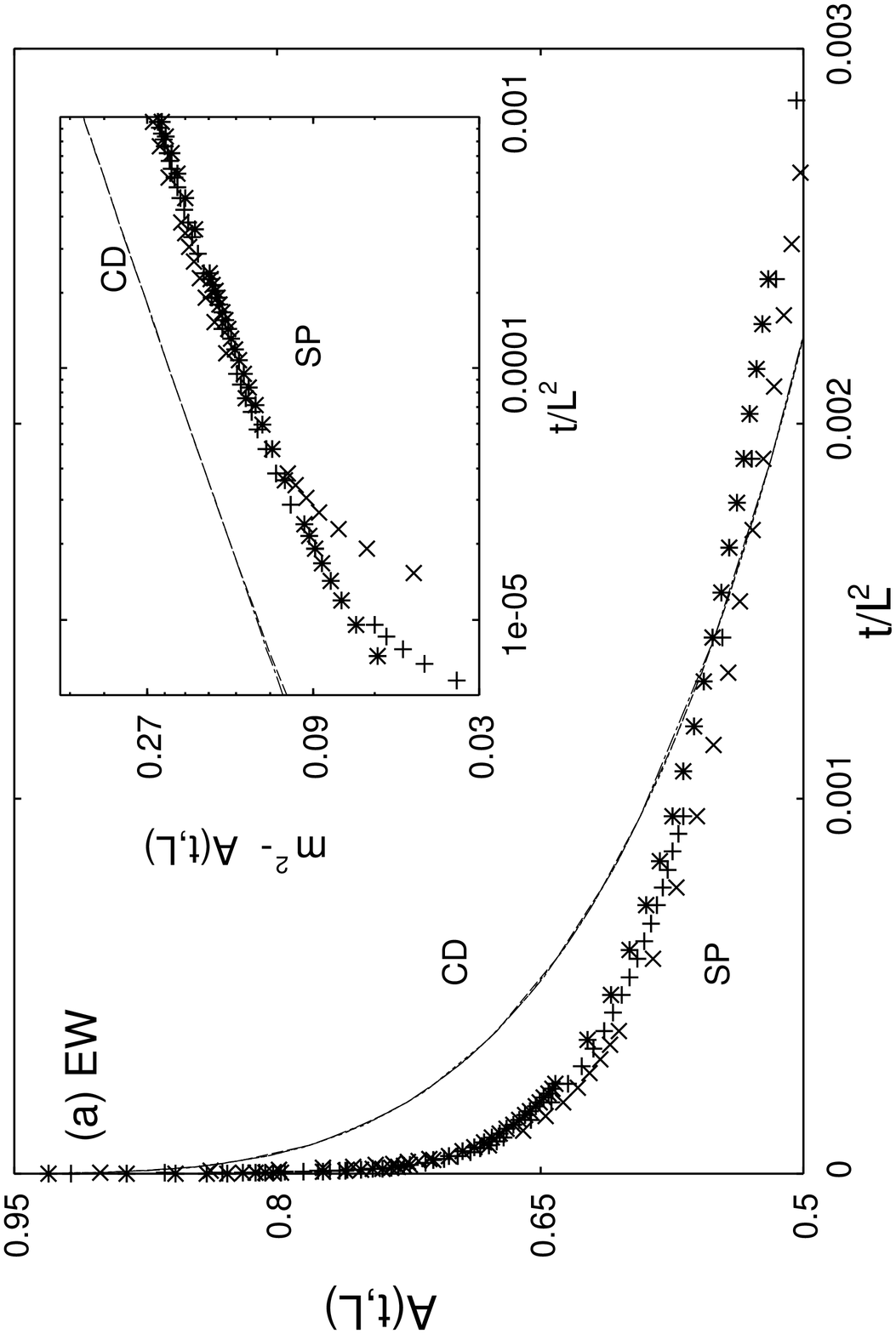}
\includegraphics[scale=0.3,angle=270]{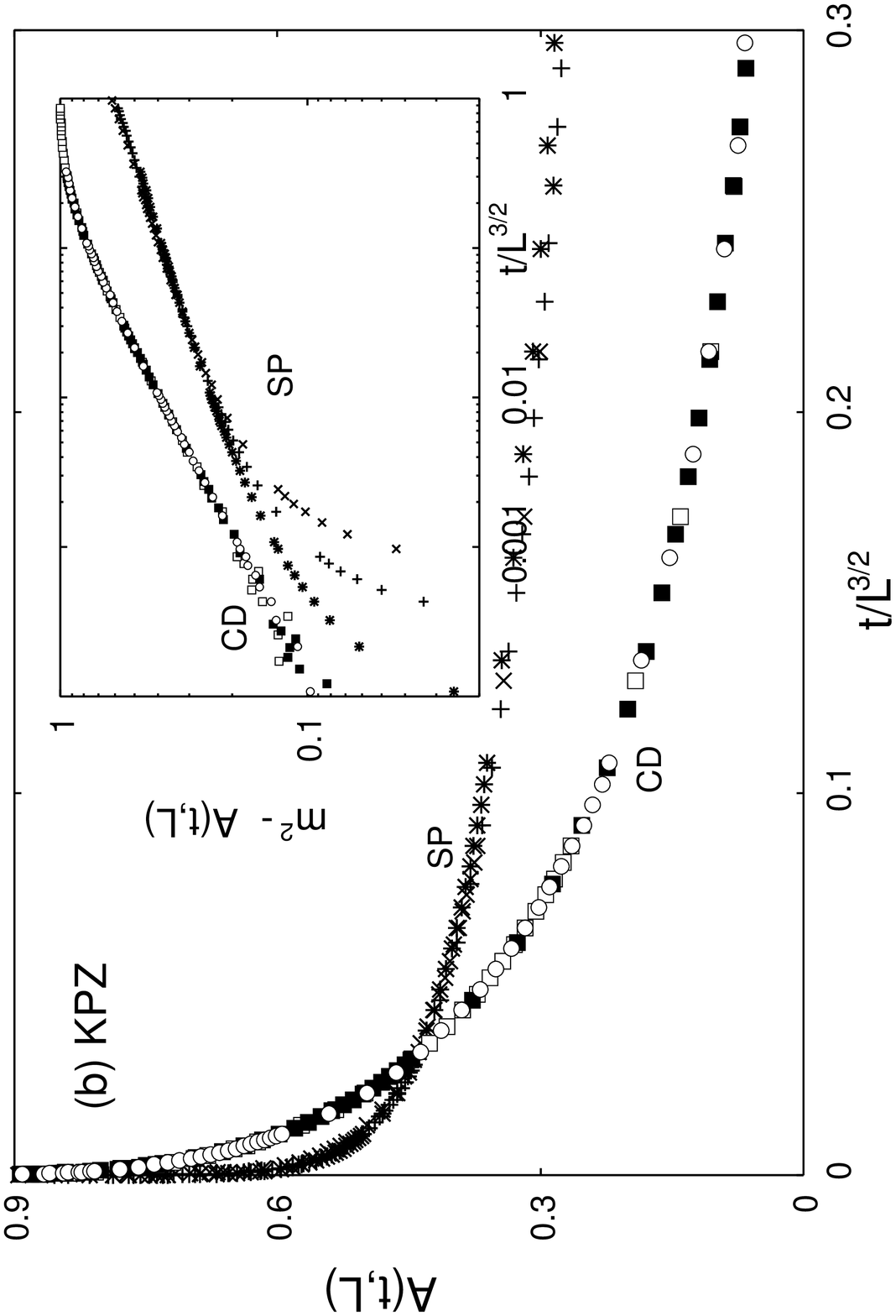}
\caption{\it Scaled auto-correlation function in steady state for SP and 
CD models for (a) EW and (b) KPZ interfaces. In both cases, we used
L=512,1024,2048. The cusp exponents were
estimated using the plots shown in the inset.}
\end{figure}
\par
For the sliding particle (SP) model, the steady state measure is not known
analytically. In our simulation, we started from a randomly disordered 
configuration and allowed a long time $\sim 10L^z$ 
for the system to reach a steady state. We then measured 
$\frac{1}{L} \sum_{i=1}^L \sigma_i (0) \sigma_i (t)$ for approximately $L^z$ 
time-steps. We waited several thousand time-steps before repeating the
procedure, and averaged over $10^4$ histories .
\par 
For particles sliding on an EW interface, we obtained a good 
scaling collapse of $A_{SP}(t,L)$ for different $L$ after rescaling time to
 $t/L^2$ [Fig.(2a)]. 
The cusp exponent was extracted by fitting $m^2-A_{SP}(t,L)$
to a power law. 
We have estimated $m^2$ by using the same technique as
 discussed in~\cite{das}. The best estimate of $m^2$ corresponds to the value
for which the structure factor has the largest power law stretch.
 We found that $m^2$ shows a systematic dependence 
on $L$ and the cusp exponent $\beta '$ is in fact quite sensitive to the value 
of $m^2$. We have used $m^2 \simeq 0.82$, our estimate from 
the largest  
system size we could access ($L=4096$). This yields $\beta ' \simeq 0.22$.
On the other hand, using 
$m_{\infty}^2 \simeq 0.85$, which we get by extrapolating the dependence
of $m^2$ on $L$ for an infinite system, we find $\beta '\simeq 0.20$. 
\par
For the SP model on a KPZ surface, we find $m^2 \simeq 0.75$. 
 Figure (2b) shows the
scaling collapse for different $L$ after rescaling the time by
$L^{3/2}$. The inset shows that 
$m^2-A_{SP}(t,L)$ follows a power law and the exponent 
is found to be $\beta ' \simeq 0.18$. The value of $\beta '$ obtained using
$m_{\infty}^2$ is $\simeq 0.17$.  
\section{Auto-correlation in Aging Regime}
The aging auto-correlation function ${\cal A}(t_1,t_2)$ is defined as
 $ \langle \sigma_i (t_1) \sigma_i (t_1+t_2) \rangle $ for the particles
and as $ \langle s_i (t_1) s_i (t_1+t_2) \rangle $ for the CD variables.
${\cal A}(t_1,t_2)$
depends on both $t_1$ and $t_2$. For $1 \ll t_1, t_2 \ll L^z$,
${\cal A}(t_1,t_2)$
is  a function of $\frac{t_1}{t_2}$, as expected 
for phase ordering systems~\cite{a.bray}. In 
the limit when
$t_2 \gg t_1$, this scaling function has a power law decay 
\begin{equation}
{\cal A}(t_1,t_2) \sim \left ( \frac{t_1}{t_2} \right ) ^{\gamma} \mbox{ for }
 t_2 \gg t_1,           \label{eq:gamma}
\end{equation}
while in the opposite limit, $t_1 \gg t_2$, the scaling function has the form
\begin{equation}
{\cal A}(t_1,t_2) \sim m^2 \left ( 1-b_1 \left ( \frac{t_2}{t_1} 
\right )
^ {\beta '} \right ) \mbox{ for } \frac{t_2}{t_1} \rightarrow 0
\label{eq:age}
\end{equation} 
This is similar to the form of the steady-state auto-correlation in 
Eq.(\ref{eq:beta}) with $L$ replaced by  $t_1 ^{1/z}$, meaning 
 that locally the system has reached steady state over a length
scale of $t_1^{1/z}$. 
\par
We first present our results on the CD model. As in the case of steady-state
 auto-correlation, we have been able to calculate ${\cal A}_{CD}(t_1,t_2)$
 for an EW
 surface analytically. Following similar steps as in the last section, 
 we obtain  
\begin{widetext}
\begin{equation}
{\cal A}_{CD}(t_1,t_2)=\frac{2}{\pi} sin^{-1} \left [ \frac{\sum_{k \neq 0} 
\frac{\exp 
\left (-c_k t_2 \right ) -\exp \left [-c_k (2t_1 + t_2) \right ] }{c_k}}
{\left \{ \sum_{k' \neq 0} \frac{1-\exp \left (-2 c_{k'} t_1 \right )}{c_{k'}}
 \right \} ^{1/2} \left \{ \sum_{k'' \neq 0} \frac{1-\exp
 \left [ -2 c_{k''} (t_1 + t_2) \right ]}{c_{k''}}  \right \}^{1/2}} 
 \right]
\label{eq:agsum}
\end{equation}
\end{widetext}
Taking the continuum limit and using $ t_1, t_2 \ll L^2 $, we obtain
\begin{equation}
{\cal A}_{CD}(t_1,t_2) = \frac{2}{\pi} sin^{-1} \left [ \frac
{ \sqrt {2 t_1 +t_2} -
\sqrt{t_2} }{\left ( 2 t_1 \right )^{1/4} \left ( 2t_1+2t_2\right )^{1/4}}
\right ]
\end{equation}
In the limit $t_2 \gg t_1$, right hand side becomes $\frac{\sqrt{2}}{\pi}
 \left ( \frac{t_1}{t_2} \right ) ^{3/4} $. Comparing with Eq.(\ref{eq:gamma}),
we get $\gamma=\frac{3}{4}$.
In the opposite limit, when  $t_1 \gg t_2$, the right hand  side becomes, after
simplification,  
\begin{equation}
{\cal A}_{CD}(t_1,t_2) \approx 1 - \frac{2^{\frac{5}{4}}}{\pi}
 \left ( \frac{t_2}{t_1} 
\right ) ^{1/4}
\end{equation}
Comparing with Eq.(\ref{eq:age}), we find $\beta' =1/4$, as expected.    
\par
Figure (3a) shows the numerical evaluation of the discrete 
sum in Eq.(\ref{eq:agsum}). The power law characterizing the decay
 has been shown in the inset.  
\begin{figure} [h]
\includegraphics[scale=0.3,angle=270]{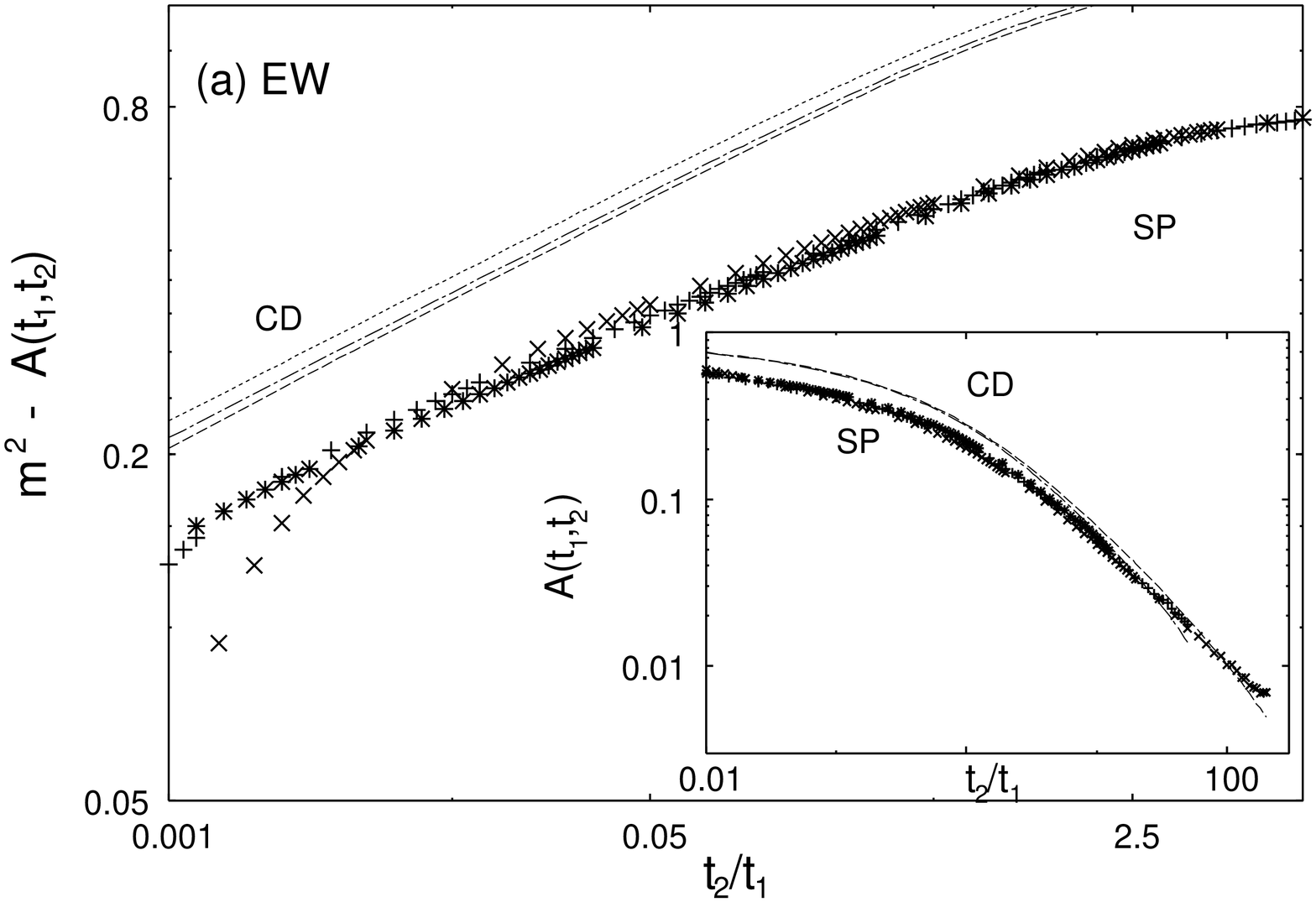}
\includegraphics[scale=0.3,angle=270]{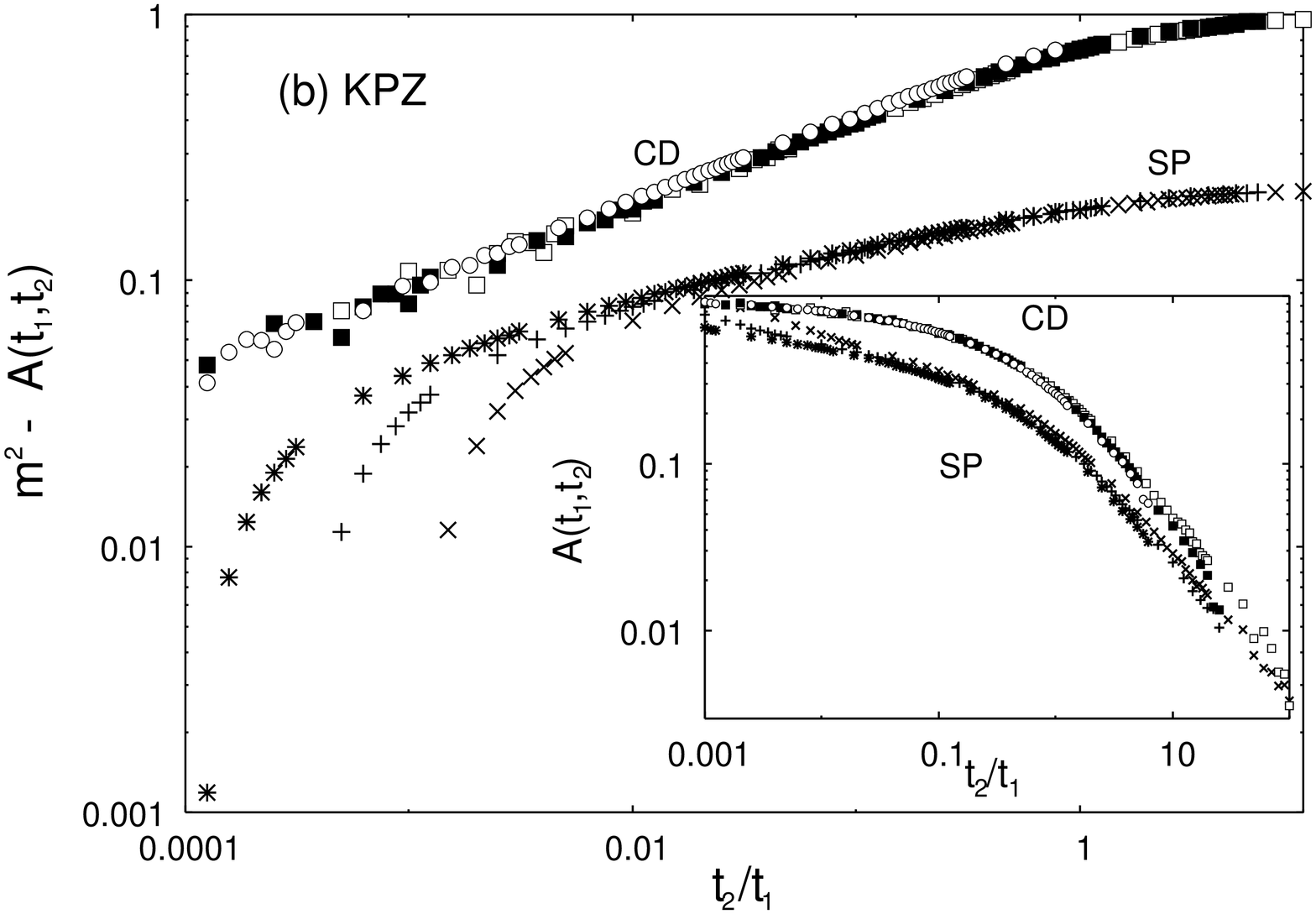}
\caption{\it Aging auto-correlation for CD and SP models with (a) EW
 and (b) KPZ interfaces. 
 The cusp exponent $\beta'$ was determined
for $t_1 \gg t_2$, after subtraction from $m^2$. The CD model data in (a)
has been multiplied by $1.5$ to distinguish
it from the SP model data points.
The inset shows the power law behavior in the regime 
$t_1 \ll t_2$. We used $L=2048$ in (a) and $L=8192$ in (b). The Inset shows the
data with $t_1=500,2000,8000$ in both (a) and (b). 
For extraction of $\beta'$, we used $t_1=2000,8000,32000$.}
\end{figure}
\par
In our Monte Carlo simulations, we have a spatial average as well as
an average over $10^4$ histories. For the CD model of a KPZ surface, we started
with a flat interface as an initial condition and evolved it in time to
measure ${\cal A}_{CD}(t_1,t_2)$. The results are shown in
fig.(3b). The best estimate of the cusp exponent corresponds to $t_1 = 32000$
and the error bar is based on its values for $t_1=2000,8000$. 
This finally gives $\beta ' =0.31\pm0.01$, which is close to the steady state value. 
The inset shows the power law decay and the 
exponent $\gamma$  takes the value $0.84\pm0.03$. Here, the best estimate
is for $t_1=500$ and the error-bar is for $t_1=2000,8000$. 
\par
For the SP model on an EW interface, we start with randomly distributed
particles on a random surface profile. 
The aging auto-correlation ${\cal A}_{SP}(t_1,t_2)$ shows a scaling collapse
as plotted against 
$t_2/t_1$ [see fig.(3a)]. The  value of the cusp exponent $\beta'$
is $0.20\pm 0.02$, close to its steady state value.
The inset shows plot in the regime $t_2 \gg t_1$. The  
 power law exponent in this case is $\gamma=0.69\pm0.02$.
\par
The SP model on a KPZ surface also starts with the random initial condition.
The data are shown in fig.(3b).  
 The exponents are
$\beta ' = 0.17\pm0.01$ and $\gamma=0.82\pm0.04$. 

\section{Space-time Correlation in Steady State}
In this section, we discuss 
 the behavior of 
space-time correlation $G(r,t,L)$ defined in
 steady state as 
 $\langle \sigma_i (0) \sigma_{i+r} (t) \rangle $ 
for the particles and $\langle s_i(0)
s_{i+r}(t) \rangle $ for the CD variable. $G(r,t,L)$ does not show any
$L-$independent scaling between $r$ and $t$. Rather, it is 
 a function of the 
scaled variables $\xi = r/L$ and $\tau = t/L^z$
\begin{equation}
G(r,t,L) = g(\xi, \tau).
\end{equation}  
With $\xi$ held fixed, $g$ shows an interesting
 non-monotonic behavior with $\tau$. $g(\xi, 0)$ reduces to the 
pair correlation function $f(\xi)$ [see
 Eq.(\ref{eq:f})], and as $\tau$ increases, $g(\xi, \tau)$ is observed to 
rise and attain
a peak [see Fig.(4a)].
 Finally for larger $\tau$, it decreases and merges with the 
auto-correlation scaling function $h(\tau)$ [see Eq.(\ref{eq:h})]. 
Note that $G(r=0,t,L) \equiv A(t,L)$, 
and by continuity, we expect that when $\xi^z \ll \tau$, the scaling
 function
should behave like $h(\tau)$.
  From
our knowledge of the scaling functions $f(\xi)$ and $h(\tau)$, we have been
able to verify that $f(\xi) < h(\tau = \xi ^z)$. This implies 
 that $g(\xi, \tau)$ must show an initial rise.
\par
 For the CD model on an EW
 interface, 
\begin{equation}
G_{CD}(r,t,L) = \frac{2}{\pi} sin^{-1} \left [ 
\frac{\sum_{k>0}\frac{\exp \left (-c_k t
 \right ) 2\cos (kr)}{c_k}}{\sum_{k>0} \frac{1}{c_k}} \right ] .
\end{equation}
We have evaluated this sum numerically and plotted it against  
$\tau$, for a fixed value of 
$\xi$ in Fig.(4a), inset which shows its non-monotonic nature.
In the continuum limit, the argument of arcsine takes the form
$\left [ 2 \cos(2 \pi \xi) - 2 \pi ^2 \xi + 2 \pi \xi Si(2 \pi \xi) -
2 \pi \nu_1 N(\xi,\tau) \right ]$  where $N(\xi, \tau)$ is defined as 
 $\int_{0}^{\tau} dy \sqrt{\frac{\pi}{\nu_1 y}} \exp\left (-\frac{\xi^2}{4\nu_1y}\right )
\left [ erf \left ( 2 \pi \sqrt{\nu_1 y}  -\frac{i\xi}{2 \sqrt{\nu_1 y}}\right ) -1 \right ] 
$
which shows explicitly that $G_{CD}(r,t,L)$ is a 
 function of $\xi$ and $\tau$ only.  
\begin{figure} [h]
\includegraphics[scale=0.3,angle=270] {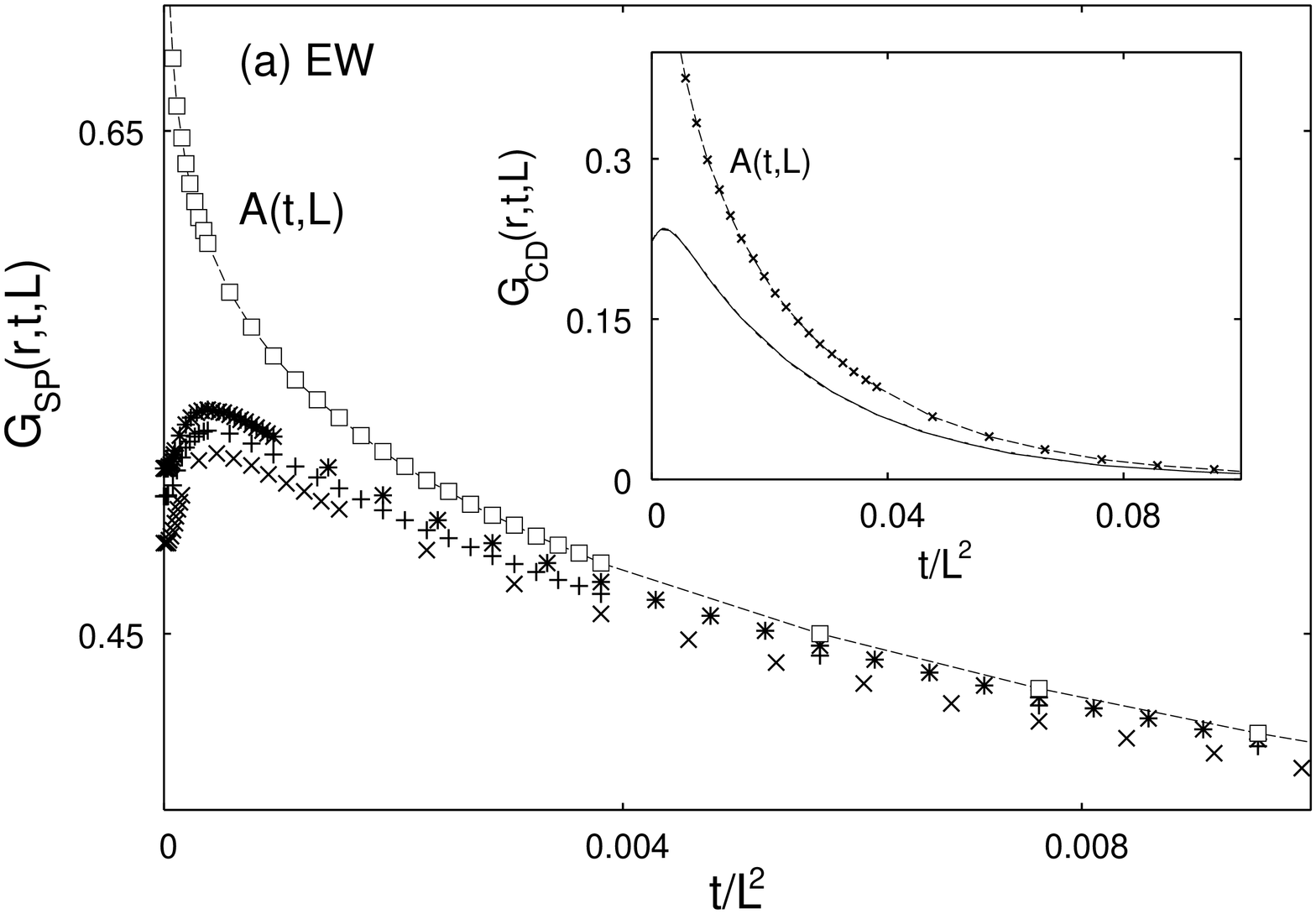}
\includegraphics[scale=0.3,angle=270]{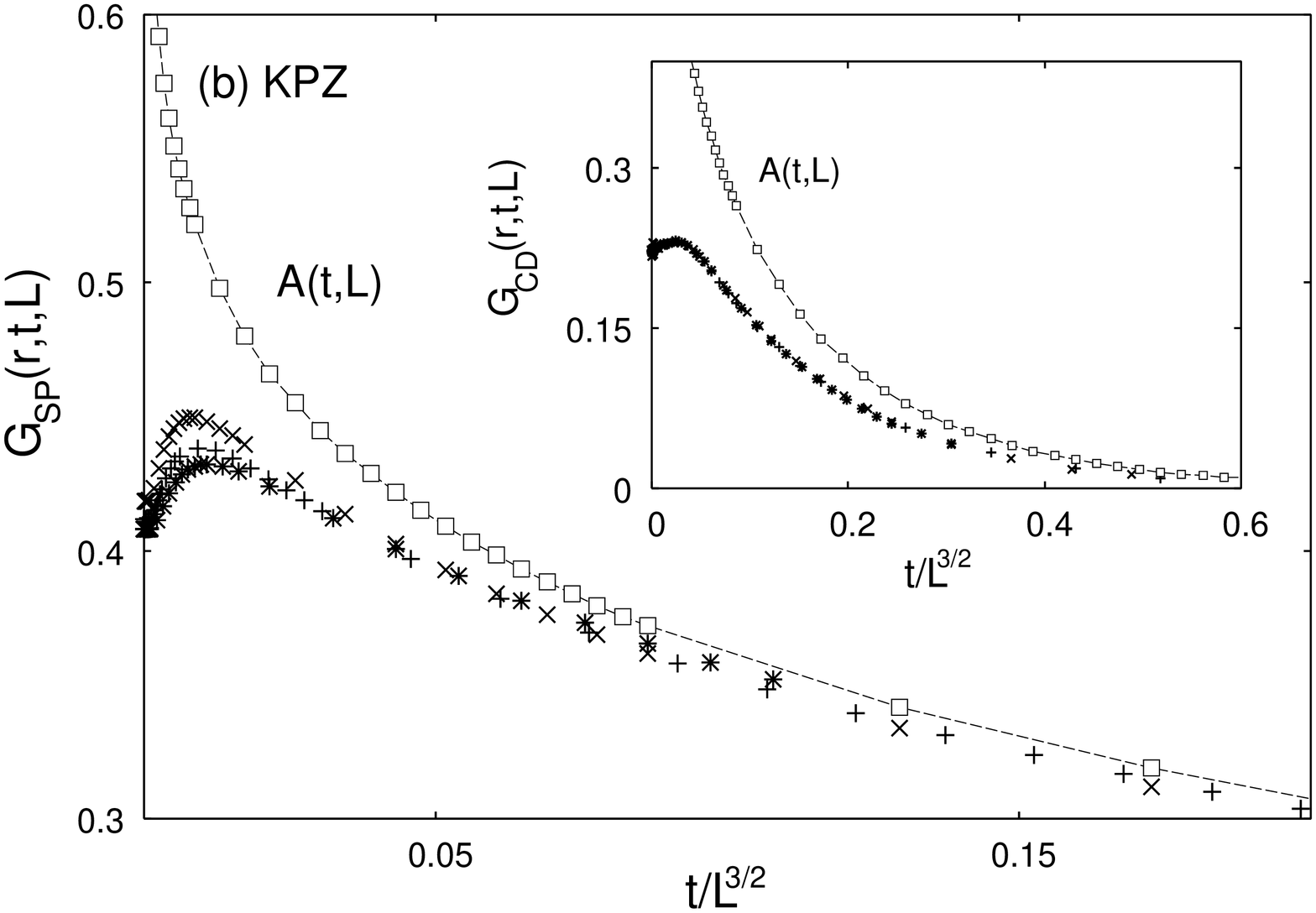}
\caption{\it The time dependence of 
$G(r,t,L)$ is shown for particles on an (a)EW and
(b) KPZ surface for
 $\frac{r}{L}=0.016$. The values of $L$ are $256,512,1024$ for (a) and
 $512,1024,2048$ for (b). The scaled 
auto-correlation is also shown, for comparison. The insets show
the same quantity calculated for the corresponding CD model with
$\frac{r}{L}=0.125$ for both cases.}
\end{figure}
\par
To measure $G_{SP}(r,t,L)$ for particles on an EW surface we performed Monte Carlo
simulations as before. After equilibrating the system, we measure 
$\frac{1}{L}\sum_{i=1}^{L} \sigma_i(0) \sigma_{i+r}(t)$ for about $L^z/10$
time steps, then after a gap of a
few hundred time steps, we take another set of data. We finally average
over $10^5$ such histories. The results are shown in Fig.(4a)   
where we have also included the scaling function $h(\tau)$ to compare the 
long time behavior. The corresponding results for KPZ surface are shown in 
Fig.(4b).
\section{Largest Cluster in Steady State}
One of the key characteristics of FDPO is the presence of strong
 fluctuations, even in the thermodynamic limit. In the steady state, large
 clusters 
are present in the system and the cluster size distribution follows a 
power law. 
As a result of fluctuations, these clusters  undergo large changes
in their lengths, associated with the fact that the macroscopic state of
 the system keeps
changing. For a system of size $L$, the typical lifetime of a macrostate 
scales as $L^z$. The question arises: if the lifetime of a state is so much
smaller than exponential,
 in what sense can we call such a state a `phase'? We have addressed this
 question in the 
following way. Let $l_{max} (t)$ be the length of the largest cluster present
in the system at time $t$. In a disordered state, this 
length scales as $\log L$.
But starting from a random initial configuration, as the system approaches 
steady state, $l_{max} (t)$, although a fluctuating quantity, shows an 
increasing trend. Finally, in steady state, $l_{max} (t)$ is still fluctuating,
thereby changing the macroscopic state of the system. But $l_{max} (t)$
 continues to remain substantially above its disordered state value $\log L$.
 In 
other words, the system manages to retain its ordered character despite
steady state fluctuations. The system continues to move from one macroscopic 
state to other over a time-scale of $L^z$. But each of these states are ordered
in the sense that they all correspond to large values of $l_{max} (t)$. 
\par
We
have studied the distribution of $l_{max}$ in steady state as well as in 
disordered state. Our studies show that as system size increases, 
the overlap
between these two distributions falls off. This means that as $L$ grows, it
is increasingly unlikely for the steady state $l_{max}$ to come down as low as
its disordered state value. The time-scale
for such a transition in fact grows exponentially with $L$. 
\par
The disordered state distribution was obtained by averaging over 
$10^8$ data points. The mean of this distribution scales as $\log L$ as
mentioned earlier.
\par
\begin{figure} [h]
\includegraphics[scale=0.3,angle=270]{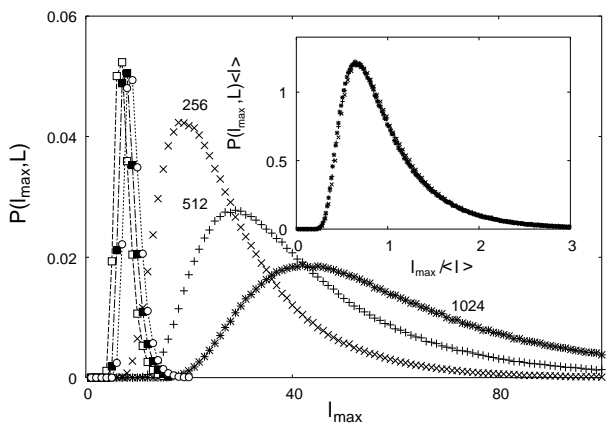}
\caption{\it The distribution of the length of the largest cluster 
$P(l_{max},L)$ for particles advected by KPZ surface is
shown for $L=256,512,1024$, with the scaling collapse in the inset.The
curves to the left show the same distribution in disordered state, after
 rescaling the y-axis by $0.2$}
\label{fig:pkplarg}
\end{figure}

After the system has reached steady state, we measure the largest
cluster present in that configuration. We let the configuration 
 evolve in time and after 
waiting for few hundred time steps, we again measure $l_{max} (t)$. We 
obtain the distribution $P(l_{max},L)$ after normalizing over $10^6$ such 
data points. As shown in the following figures, $P(l_{max},L)$ for 
different $L$ values undergo a scaling collapse as $l_{max}$ is rescaled
by the mean of the distribution $\langle l \rangle$. We have found
that $\langle l \rangle \sim L^{\phi}$, where the exponent $\phi$ depends 
on the dynamical rules. For particles on an EW surface $\phi \simeq 0.86$,
whereas for
KPZ advection,  $\phi \simeq 0.60$ while for KPZ anti-advection
$\phi \simeq 0.91$. We show the data for KPZ advection in
fig.(\ref{fig:pkplarg}).

\section{Discussion}
In this paper, we have studied the dynamics of interacting 
 passive scalars driven by a
 fluctuating Edwards-Wilkinson or Kardar-Parisi-Zhang
 surface (or equivalently, a Burgers fluid), by characterizing the 
 scaling properties of 
 correlation functions in steady state and those of aging correlations
 during the approach  to steady state. It is instructive to compare our
 results with earlier work on the dynamics of passive scalars in different
 contexts. 
 \par
 Mitra and Pandit~\cite{mitra} studied the dynamical properties of a system of
 non-interacting passive particles, advected by an incompressible 
 fluid, whose velocity field is drawn from the Kraichnan ensemble, 
 and therefore
 has power law correlations in space, but is delta-correlated in
 time. By contrast, we have studied passive particles with hard-core
 interactions, advected (in the Burgers case) by a compressible flow which has
 power law correlations in time. The compressible nature of the flow results
 in particles being driven together in our case. The resulting state 
 is described by a
 space-time correlation function $G(r,t,L)$ which is a function of the
 scaling
 combinations $r/L$ and $t/L^z$, as in any phase-ordering system. However, 
 there is no
 non-trivial  $r-t$
 scaling in the limit $L \rightarrow \infty$. 
 On the contrary, in~\cite{mitra}, $G(r,t,L)$ was found to show a
scaling between $r$ and $t$, with $t \sim r^z$ for fixed $L$. 
This difference of behavior is presumably a
reflection of the strong differences between passive scalars with 
clustering or phase-ordering tendencies, and those which spread out in space. 
\par
Even when the driving fluid is compressible, the degree of 
clustering of the
passive particles depends on the nature of interactions between them.  
In the presence of  hard-core interactions, the system reaches a phase-ordered 
state, albeit one with strong fluctuations.
As a consequence, in the limit of small scaling argument,
 the spatial and temporal
 correlation functions show a cuspy approach to a finite intercept.
However, in the absence of any interaction, the
passive particles go into a more strongly clustered state, where 
the correlation functions show a power law divergence at
the origin~\cite{nagar}. 
\par
Finally, the study of the largest cluster allows us to arrive at a simple 
picture of a fluctuation dominated
phase-ordered state. Despite  
 the presence of strong fluctuations, the system never loses 
its ordered character.  Fluctuations
carry the system from one ordered configuration to another macroscopically
distinct one, over a time-scale
 $\sim L^z$. However, the probability for the system to
leave this attractor of ordered states vanishes exponentially
with the system size.
\section{Acknowledgments}
We acknowledge useful discussions with S.N. Majumdar and S.Ramaswamy. 
SC would like to thank
TIFR Endowment Fund for partial financial support.


\begin{thebibliography}{99}
\bibitem{siggia} B.I.Shraiman and E.D.Siggia {\it Nature} {\bf 405}, 
369 (2000).
\bibitem{falkovich} E.Balkovsky, G.Falkovich and A.Fouxon {\it Phys. Rev. Lett.}
{\bf 86}, 2790 (2000).
\bibitem{vergassola} K.Gawedzki and M.Vergassola {\it Physica D} {\bf 138}, 63 
(2000).
\bibitem{medina} E.Medina, T.Hwa, M.Kardar, Y.-C.Zhang {\it Phys. Rev. A}
{\bf 39}, 3053 (1989).
\bibitem{nagar} A.Nagar, M.Barma and S.N.Majumdar {\it Phys. Rev. Lett}
 {\bf 94}, 240601 (2005).
 \bibitem{drossel} B.Drossel and M.Kardar {\it Phys. Rev. B} {\bf 66}, 
 195414 (2002)
\bibitem{das} D.Das, M.Barma and S.N.Majumdar {\it Phys. Rev. E} {\bf 64},
 046126 (2001).
 \bibitem{manoj} G.Manoj and M.Barma {\it J. Stat. Phys.} {\bf 110},
 1305 (2003).
\bibitem{mitra} D.Mitra and R.Pandit {\it nlin.CD/0412013}. 
\bibitem{ew} S.F.Edwards, D.R.Wilkinson {\it Proc. R. Soc. London A}
{\bf 381}, 17 (1982).
\bibitem{barabasi} A.-L.Barab\'{a}si and H.E.Stanley, {\it Fractal Concepts
in Surface Growth} (Cambridge University Press, New York,1995).
\bibitem{a.bray} A.J.Bray {\it Adv. Phys.} {\bf 43}, 357 (1994).
\bibitem{porod} G.Porod, in {\it Small Angle X-ray Scattering},
 edited by O.Glatter and L.Kratky (Academic Press, New York, 1983).
\bibitem{meakin} P.Meakin, {\it Fractals,Scaling and Growth Far from
 Equilibrium} (Cambridge University Press, Cambridge,1998)
\end{thebibliography}
\end{document}